# FABRICATION OF SWTICHES ON POLYMER-BASED BY HOT EMBOSSING

*Chao-Heng Chien, Hui-Min Yu,*

Mechanical Engineering Department, Tatung University
40 Chung Shan N. Rd. Sec. 3 Taipei, Taiwan

## ABSTRACT

In MEMS technology, most of the devices are fabricated on glass or silicon substrate. However, this research presents a novel manufacture method that is derived from conventional hot embossing technology to fabricate the electrostatic switches on polymer material. The procedures of fabrication involve the metal deposition, photolithography, electroplating, hot embossing and hot embed techniques. The fundamental concept of the hot embed technology is that the temperature should be increased above Tg of polymer, and the polymer becomes plastic and viscous and could be molded. According to the fundamental concept, the metal layer on the silicon/glass substrate could be embedded into polymer material during the hot embossing process. Afterward, the metal layer is bonded together with the polymer after removing the substrate in the de-embossing step. Finally, the electrostatic switch is fabricated on polymethylmethacrylate(PMMA) material to demonstrate the novel method.

## 1. INTRODUCTION

Hot embossing has been a mature technology in traditional mechanical fabrication. It uses widely in various manufactures, for example, mold replication, compact disk and package. The basic principle of these techniques is the replication of a mold tool (also call a replication master), which represents the negative (inverse) structure of the desired polymer structure [1]. The fabrication process is that the temperature above the Tg (glass transition temperature) of polymer, the polymer became viscous and can be molded. After the temperature bellowing the Tg, the mold is to de-embossing and to complete the replication process [2]. Fig. 1 shows the traditional hot embossing process.

Since the MEMS technology has been developed over the past decade. More and more researches are published about the MEMS to combine with the fabrication of micro mold. Thus, it makes the hot embossing plays an important micromachining technology as microchannel [3], micropump, Lab-on-a-chip, and microlens [4]. Besides, the polymer materials are used in the micromachining because they are suitable for such mass replication technologies and to offer various advantages and properties, such as, bio-compatible, transparent, low-cost and manufacture easily. Thus, the polymer materials are popular to be used in MEMS.

The electrostatic switch is fabricated in this research. In general, the switch is constructed using the standard MEMS technologies, such as, pattern, deposition, and removing the sacrificial layer on glass/silicon substrate. In the procedure, removing the sacrificial layer is not easy to forms the gap if the electrodes have a large area. In order to solve the problem, this research provides a novel fabrication technology which is derived from the traditional hot embossing. The novel method eliminates the step of removing the sacrificial layer. Besides, the substrate is the polymer material to replace the silicon or glass. The polymer is very cheap than silicon/glass so that the cost can be saved much more.

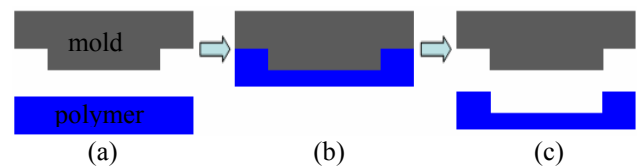

Fig. 1. The traditional hot embossing processes. (a) The mold and polymer material. (b) The hot press step. (c) To de-embossing.

## 2. DESIGN CONCEPT

Fig. 2 illustrates the schematic structure of the electrostatic switch. The electrostatic force is employed to flexure the actuating electrode forward the fixed electrode by the DC power. The switch is ON situation when the cantilever beam touches the electrode. When a potential difference is maintained between the two electrodes, the magnitude of the electrostatic force acting on the actuating electrode is given by [5]

$$F_e = \varepsilon \frac{AV^2}{2g^2} \qquad 1$$

where $\varepsilon$ is the permittivity of the air in the gap, A is the area of the fixed electrode, V is the applied voltage, and g is the gap between the two electrodes. Fig. 3 shows the 3-D structure of the electrode switch.

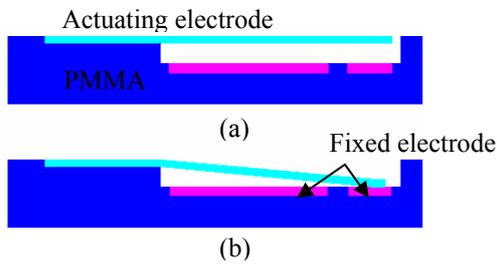

Fig. 2. The schematic structure of the electrostatic switch.

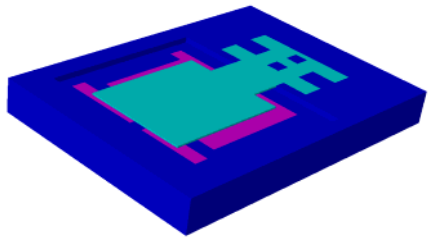

Fig. 3. The three-dimension structure of the electrostatic switch.

## 3. HOT EMBOSSING PROCESS

Fig. 4 shows the novel process to fabricate the switch on PMMA substrate by hot embossing in this research. The stamp material of metal layer is patterned and electroplated on silicon/glass wafer to become the mold for hot embossing process. At first, both of the mold and polymer material are placed on the upper and down plates of the hot embossing machine, separately, shown in Fig. 4(b). Afterward, the mold and polymer are heated and applied the clamping force to press the metal layer into the polymer for a few seconds as the temperature above the Tg of polymer, shown in Fig. 4(c). The operation temperature and clamping force are very important in this step. Because the thickness of metal layer on the silicon/glass wafer is only 10~40 $\mu$m. The large clamping force causes the disappearance of the gap and the deformation of the switch. In the last step is de-embossing and removing the silicon/glass wafer to embed the stamp material of metal layer into the polymer material, shown in Fig. 4(d). The result shows that the adhesion between the metal layer and polymer material is more than the metal layer and silicon wafer after hot embossing. Using the novel method, the multi-metal layers can be embedded into the polymer material, such as the electrostatic switch in this research. The basic physical properties of polymer materials data, polycarbonate (PC) and polymethylmethacrylate (PMMA) shown in table 1 [2,6].

Table 1 The polymer materials, PMMA and PC.

| Polymer material | Density ($10^3$kg/m$^3$) | Tg (°C) | CTE (ppm/°C) | Thermal conductivity (W/m•K) |
|---|---|---|---|---|
| PMMA | 1.17-1.2 | 85-105 | 50-90 | 0.186 |
| PC | 1.2 | 150 | 68 | 0.21 |

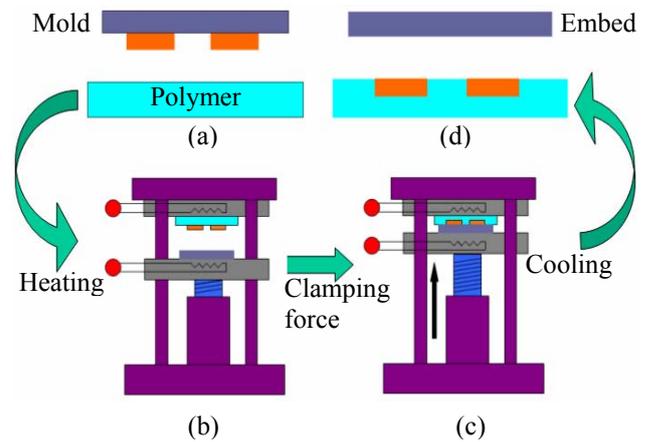

Fig. 4. The hot embossing process in this research. (a) The polymer material and mold. (b) The polymer and mold are mounted on the two plates. (c) The clamping force is applied and hold after the temperature above the Tg of polymer. (d) The stamp material of metal layer is embedded into the polymer after de-embossing.

## 4. MOLD FABRICATION

The fabrication of the electrostatic switch needed three molds, fixed electrode mold, actuating electrode mold, and gap layer mold. Fig. 5 shows the major fabrication processes of the three molds.

In the fixed electrode mold procedure, the silicon wafer with the oxidation layer is employed as the mold substrate. The solution acetone and methyl alcohol are needed to clean wafer in ultrasonic machine. The 100nm thickness of copper metal is deposited as the seed layer on the surface, then it is patterned and electroplated 20 $\mu$m thick copper on the seed layer as the electrode, shown in Fig. 5(a)~(c). Finally, the negative PR is stripped to complete the fixed electrode mode, shown in Fig. 5(d). The actual fixed actuating electrode is shown in Fig. 6.

The fabrication procedure of the actuating electrode mold is similar to the fixed electrode mold. The procedure is shown in Fig. 7. The copper electroplating layer of 12 $\mu$m thickness is patterned as being the actuating electrode, shown in Fig. 7(a)~(c). In Fig. 7(d)~(e) processes, the alignment key is made in order to align between the molds in the hot embossing process.

The purpose of the gap layer mold is to produce a gap between electrodes. Thus, the Ni metal layer is not striped from the wafer. For the purpose, the titanium is deposited to enhance the adhesion between the glass wafer and copper seed layer in order to avoid 40 $\mu$m thickness of nickel layer embed into the polymer substrate during the de-embossing process. Fig. 9 shows the fabrication procedure. The process of the actuating mold and the gap layer mold are the same. The Ni layer displaces the copper layer on the glass wafer. Fig. 10 shows the actual gap layer mold.

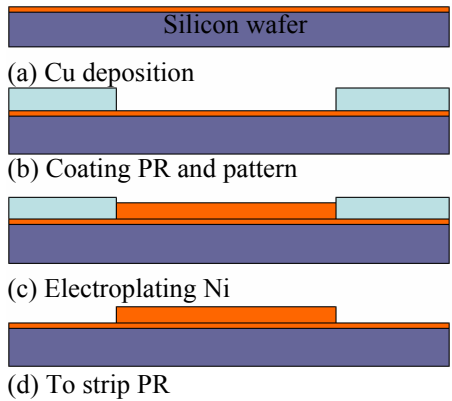

Fig. 5. The fabrication process of the fixed electrode.

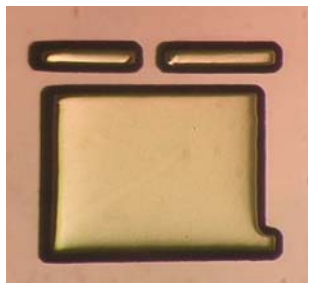

Fig. 6. The fixed electrode mold.

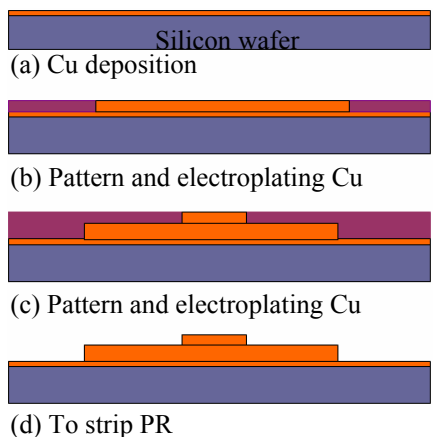

Fig. 7. The fabrication of the actuating electrode mold.

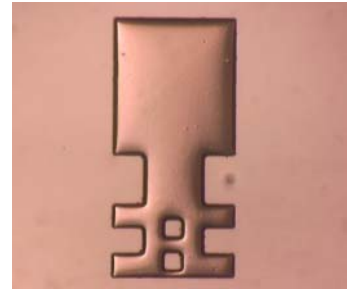

Fig. 8. The actuating electrode mold.

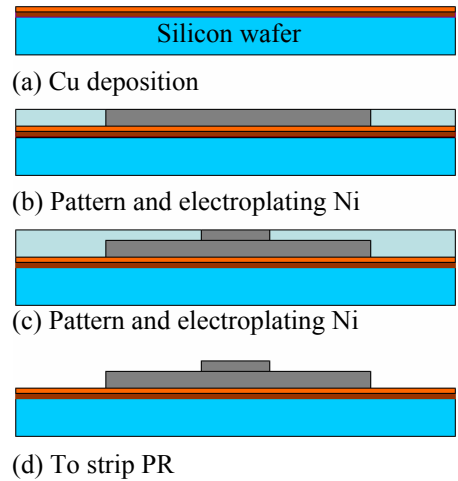

Fig. 9. The fabrication of the gap layer mold.

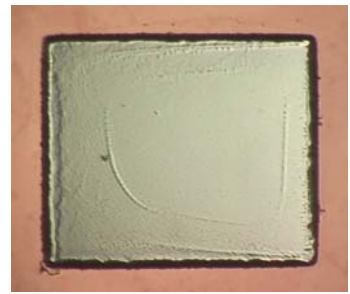

Fig. 10. The gap layer mold. The dimension is 1.15x1.15mm$^2$.

### 5. EXPERIMENT RESULTS AND DISCUSSIONS

Fig. 11 shows that the sequence of the three molds is utilized during the hot embossing process. The two stamp materials, copper, are embedded into the PMMA substrate and to be as the fixed bottom and upper electrode with the 28 $\mu$m depth of gap by hot embossing technology. The experimental results of the hot embossing are shown in Fig. 12. The result exhibits the fixed electrode to embed successfully into the PMMA and it has no protrusion in the de-embossing. The oxidation of the electrodes arises in the hot embossing process due to the temperature effect in our experiment. However, the oxidation can be prevented if the hot embossing stage in a vacuum environment. The alignment is anther important issue because the polymer materials have a large coefficient of expansion. Thus, the error will be existence easily in the multi-metal layers hot embossing process.

The clamping force has to be precise in the embedded step. Although too much clamping force can also cause the electrode embedding into the PMMA, the error of alignment and the deformation of switch are arose at the same time. The disappearance of the gap between the fixed electrode and the actuating electrode is occurred, simultaneously. On the contrary, the electrodes are not embedded into the PMMA in too small clamping force. Table 2 shows the experimental parameters of the three molds in the hot embossing process. In order to make sure the electrode is not stripped from the PMMA, the de-embossing temperature is less than the traditional hot embossing, such as fixed electrode and actuating electrode.

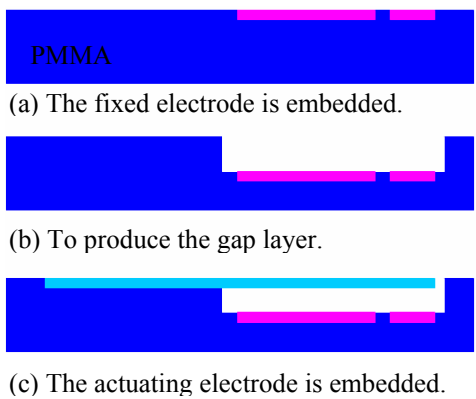

(a) The fixed electrode is embedded.

(b) To produce the gap layer.

(c) The actuating electrode is embedded.

Fig. 11. The fabrication steps of the electrostatic switch in the hot embossing process. The sequence of the molds is (a) The fixed electrode mold. (b) The gap layer electrode mold. (c) The actuating electrode mold.

Table 2. The parameters of hot embossing

|  | Fixed electrode mold | Gap layer electrode mold | Actuating electrode mold |
|---|---|---|---|
| Initial temperature (°C) | 30 | 30 | 30 |
| Embossing temperature (°C) | 120 | 120 | 120 |
| Clamping force (N/mm$^2$) | 4.4 | 2.2 | 0.1 |
| Hold time (s) | 90 | 90 | 90 |
| De-embossing temperature (°C) | 50 | 75 | 50 |

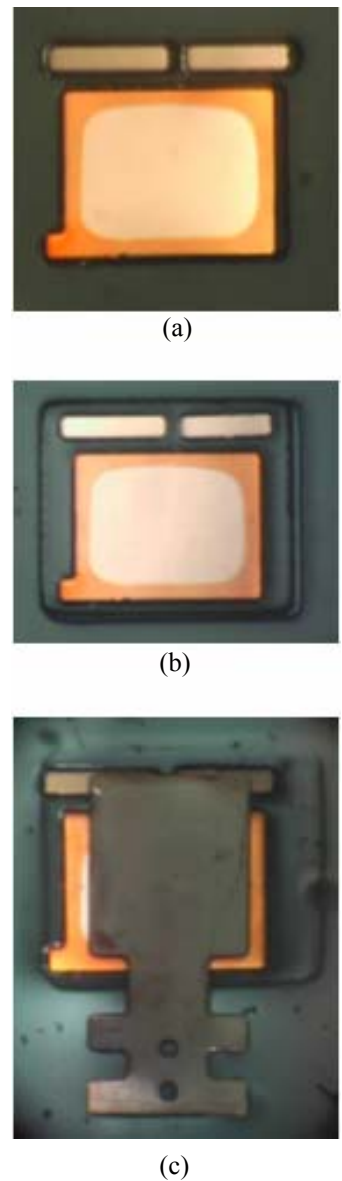

(a)

(b)

(c)

Fig. 12. The results of the hot embossing. (a) The Cu electrode is embedded into the PMMA. (b) To produce the gap layer. (c) The actuating electrode is embedded.

## 6. CONCLUSIONS

In summary, the novel fabrication method that is derived from the traditional hot embossing technology is demonstrated by way of the manufacture of the electrostatic switch. On the other hand, the novel method can be utilized to fabricate the multi-metal layers on polymer materials to replace the silicon or glass substrates. The technology makes the traditional hot embossing not only the replication tool but also a way to fabricate the active/passive device on difference substrate. The advantages are that the polymer substrates are very cheap, transparent and bio-compatible. Therefore, the devices are suitable for the application of optical and bioMEMS. In future, a precise alignment method and clamping force are needed to fabricate a micro-size device.